# Fabrication of Germanium-on-insulator in a Ge wafer with a crystalline Ge top layer and buried GeO$_2$ layer by Oxygen ion implantation

Vishal Kumar Aggarwal[1], Ankita Ghatak[1], Dinakar Kanjilal[2], Debdulal Kabiraj[2], Achintya Singha[3], Sandip Bysakh[4], Samar Kumar Medda[4], Supriya Chakraborty[5] and A. K. Raychaudhuri[4#$]

[1]Department of Condensed Matter Physics and Material Sciences, S. N. Bose National Centre for Basic Sciences, JD Block, Sector-III, Kolkata 700106, India
[2]Inter University Accelerator Centre, Aruna Asaf Ali Marg, New Delhi-110067, India
[3]Department of Physics, Bose Institute, 93/1, Acharya Prafulla Chandra Road, Kolkata 700009, India
[4]CSIR-Central Glass and Ceramic Research Institute, 196 Raja S C Mullick Road, Kolkata 700032, India
[5]Indian Association for the Cultivation of Science, 2A and 2B Raja S C Mullick Road, Kolkata 700032, India

[#]Email: arupraychaudhuri4217@gmail.com, Corresponding Author.

[$]Orchid ID: 0000-0003-1229-9532

## Abstract

The paper reports fabrication of Germanium-on-Insulator (GeOI) wafer by Oxygen ion implantation of an undoped single crystalline Ge wafer of orientation (100). O$^+$ ions of energy 200 keV were implanted to a fluence of 1.9 x 10$^{18}$ ions-cm$^{-2}$ and the implanted wafer was subjected to Rapid Thermal Annealing to 650$^0$ C. The resulting wafer has a topcrystalline Ge layer of ~ 220 nm thickness and Buried Oxide layer (BOX) layer of good quality crystalline GeO$_2$ with thickness around 0.75μm. The crystalline GeO$_2$ layer has hexagonal crystal structure with lattice constants close to the standard values. Raman Spectroscopy, cross-sectional HRTEM with SAED and EDS established that the top Ge layer was recrystallized during annealing with faceted crystallites. The top layer (resistivity ≈32 ohm.cm) has a small tensile strain of around +0.4% and has estimated dislocation density of 2.7x10$^7$cm$^{-2}$. The thickness, crystallinity and electrical characteristics of the top layer and thequality of the BOX layer of GeO$_2$ are such that it can be utilized for device fabrication.

# 1. Introduction

Semiconductor materials like Silicon-on-Insulator (SOI) are increasingly being used for fabricating electronic devices [1]. The top crystalline semiconducting surface can be tailored and is used for microelectronic applications. The buried oxide layer (BOX) which for SOI is $SiO_2$ the dielectric layer that separates the top device layer and the handler wafer. These substrates are used for opto-electronic applications as well [2,3]. Recently high sensitivity broad band optical-detectors (UV to NIR) have been produced by fabricating micro and sub-micron width arrays on SOI wafers [4,5].

In this paper we address issues related to Ge-on-Insulator (GeOI) which has the potentialto become an important enabling material for future semiconductor industry. GeOI has almost all the technical capabilities that are observed in SOI [6]. In addition, Ge has high carrier mobility, high carrier concentration and lower effective mass than that of Si and relatively small band gap. These characteristics make Ge an attractive electronic and opto-electronic material and it has certain advantages over Si. Recently attention has been directed to GeOI and Ge on SOI (Ge grown on top Si layer on SOI) which are emerging as important components for integrated Mid-IR applications beyond $2 \mu m$ [7]. Two-dimensional pure Ge photonic crystals have been reported on a GeOI substrate [8]. Recently, the junction-less GeOI pMOSFETs have been demonstrated with planar MOSFET and FinFET structures have been fabricated on GeOI substrates. Superior electrical properties have been realized for these devices, thanks to the high hole mobility in the channels [9-11]. While SOI wafers (mostly made by wafer bonding of two $Si/SiO_2$ wafers) are commonly available commercially from a number of sources, the availability of GeOI wafers is not that common. Even if it is available it is costly. This is primarily because of the difficulty in fabricating such a wafer [6]. GeOI wafer can be fabricated by Ge condensation technique where a thin Ge layer is made on Si that is supported on a BOX [12]. It is also made by Liquid Phase Epitaxy [13]. Recently GeOI wafers of larger dimensions are being made by epitaxy, bonding and layer transfer [14,15]. In this method a Ge layer is grown on Si (epilayer). Atomic Layer Deposition (ALD) grown $Al_2O_3$ is then used to coverthis layer. Another Si wafer also covered with $Al_2O_3$ is then bonded with the Ge containing wafer by bonding through $Al_2O_3$ which serves as the BOX layer. The Si layer supporting the Geis etched out. This method allows a low dislocation density (DD) ~ $10^6$ $cm^{-2}$. This is much better than the DD > $10^7$ $cm^{-2}$ seen in other GeOI wafers.

A technique called Smart Cut$^{TM}$ has also been used to make larger wafers with high quality top Ge of different thicknesseson a BOX layer of $SiO_2$. It is fabricated by a process that uses bonding of $SiO_2$grown on the handler Si wafer and the active Ge wafer. $H^+$ implantation is made on the $SiO_2$ layer on Ge that helps in separating this layer [16-18]. In case of SOI, creation of a buried oxide of $SiO_2$ as an insulating isolation layer using energetic Oxygen ion implantation is well developed and is known as SIMOX (Separation by Implanted Oxygen) process [19,20]. In case of Germanium based structures like GeOI there are not too many investigations done on the SIMOX type process. In case of Germanium most of the efforts were on SGOI (SiGe-on-insulators) where the oxygen had been implanted on to the backing Si layer of a Si-Ge /Si heterostructures [21,22]. This utilizes similar processes as that used for making SOI wafers by SIMOX.

It has been shown long back that implantation of 200keV Oxygen ions in pure Ge (not alloyed with Si) with a fluence of $3 \times 10^{17}$ $cm^{-2}$ and post deposition annealing at $500^0$ C can lead to formation of $GeO_x$ with x ≈2 as confirmed by IR spectroscopy. From the IR absorption it was confirmed that the $GeO_2$ so formed is 4-fold coordinated (like $\beta$ – Quartz) [23]. Subsequent investigations using oxygen ions to 180 keV at fluence upto $2 \times 10^{18}$ cm$^{-2}$ and energy of and post implantation annealing to high temperature (upto 1000$^0$C) showed formation of buried GeO$_x$ with x ≈1.4, which was interpreted as a mixture of GeO$_2$ and GeO [24]. An implantation study on single crystal Ge wafer using higher beam current (with lower energy of 45keV) created a buried layer of GeO$_2$ at an average depth of 55nm for fluence exceeding $1.5 \times 10^{18}$ cm$^{-2}$. Redistribution of Oxygen at larger depth has been observed [25]. These early investigations though established that a buried layer of GeO$_x$ with x ≈ 2 can be formed, they did not address the important question whether a top Ge layer of sufficient crystalline quality and good electrical mobility could be created so that a reasonable GeOI wafer could be made.

In this investigation we would like to have a fresh look at the process of fabrication of GeOI by ion implantation and revisit the problem. In addition to the scientific curiosity, the motivation for a fresh look at the implantation process to fabricate GeOI is to explore whether one can utilize the ease with which the ion implantation based process can be carried out compared to other methods. This process is also advantageous where large wafer size may not be needed. We investigate, in particular, the important question whether the process of energetic Oxygen implantation in Ge (and subsequent annealing) does indeed create a top crystalline Ge layer on the buried oxide of sufficient thickness. Using Raman Spectroscopy, Grazing Incidence X-Ray diffraction (GIXRD), High resolution Transmission Electron Microscopy (HRTEM) on cross-sectional samples and electrical resistance measurements, we show that with Oxygen ion implantation (200keV and fluence $\sim 2 \times 10^{18}$ions-cm$^{-2}$) and post-implantation Rapid Thermal Annealing (RTA), a top (oxygen free) crystalline Ge layer of thickness ≥ 200nm can be created with dislocation density of $\sim 2 \times 10^7$ cm$^{-2}$ and a tensile strain of ~ 0.4%-0.5%. Existence of such a top Ge layer on such a buried GeO$_2$ created by Oxygen ion implantation has not been reported before.

## 2. Experimental Details

The experiment was done on a single crystal (nominally undoped) Germanium wafer (1 cm x 1 cm x thickness 0.5mm) of orientation <100>. Pristine wafer having an average mobility of ~1130 ± 35 cm$^2$/V-s, carrier concentration of = (1.25 ± 0.04) x10$^{14}$ cm$^{-3}$ and resistivity of 44 Ohm-cm has been used. This was used as a sample for implantation. Figure 1 shows the schematic of the process used to make the GeOI. The wafer was implanted by uniformly scanned beam of 200kev Oxygen ions from an implanter at a fluence of 2x10$^{18}$ ions-cm$^{-2}$. After implantation, the sample was subjected to Rapid Thermal Annealing (RTA) to restore the crystallinity of the implanted embedded oxide and also to recrystallize the top Ge layer.

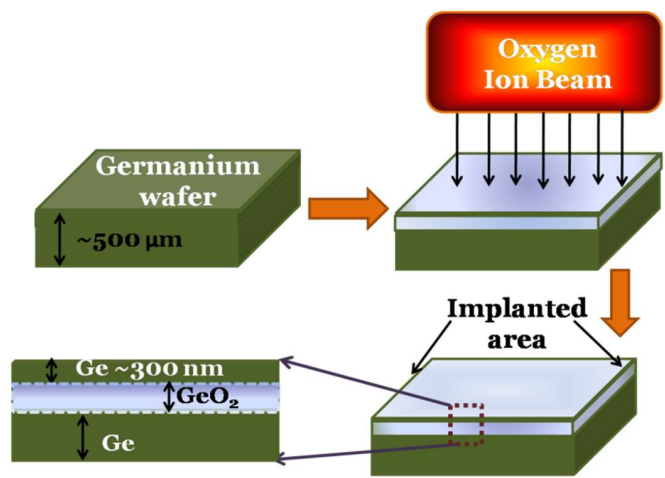

Fig. 1. Schematic of 200keV O$^+$ ion implantation.

Rapid thermal annealing (RTA) of as-irradiated wafers was carried out in inert atmosphere (Nitrogen) at 650°C at a rate of 20 °C/sec with holding time of 3 minutes to get the top Ge layer reconstructed. As a check of the effect of prolonged annealing, the irradiated wafer was annealed in Nitrogen ambient at 650°C for 1 hour at a rate of 5 °C/min. The temperature of $650^0$ C is more than the glass transition temperature ($\approx 580^0$C) of $GeO_2$ [26]. The annealing temperature thus ensures that the amorphous $GeO_2$ layer produced by ion implantation turns to crystalline during the annealing process.

## 2.1. *Structural, compositional and electrical characterization methods:*

The wafers were structurally characterized by conventional X-ray diffraction (XRD). Post implantation and RTA, there crystallized top Ge layer in the GeOI so formed was characterized by Rocking curve using grazing incidence angle of $0.3^0$ in a Rigaku Smart Lab X-ray diffractometer (GIXRD) operating at 9 kW (200 mA, 45 kV) using Cu-Ka ($\lambda$ =1.54059 Å).

Raman Spectroscopy was done at different stages of the implantation and annealing to monitor the extent of crystallinity of the Ge layer at the top. A LAB RAM HR Raman spectrometer coupled with Peltier device cooled CCD was used to collect the data counts for the particular wafer. The excitation from the LASER was focused with 100X objective lens and the spot has a Gaussian profile. The FWHM of 488 nm laser is ~700 nm and for 785 nm laser it is ~1100 nm. Two excitation wavelengths at 488nm and 785nm were used to add a depth perspective to the Raman data.

Transmission Electron Microscopy (TEM) and High Resolution TEM (HRTEM) along with Energy Dispersive X-ray Spectroscopy (EDS) were carried out to characterize the samples post-implantation and post-RTA for structural and compositional analysis. The experiments were done on cross-sectional samples. Scanning TEM-EDS spectrum imaging technique was used for elemental mapping or line-scan profiling using FEI-TIA™ application software. In the EDS spectra, Ge-K and O-K X-ray lines were used for quantification and narrow individual energy windows covering these characteristic X-ray lines were selected for mapping of the respective elements by the spectrum imaging technique. Due to limited number characteristic X-ray lines used, the Ge and O composition mapping (i.e, the wt %) by EDS will be semi-quantitative. The EDS spectrum imaging data for elemental mapping were collected over 120x120 pixel area to cover 1500x1500 nm square area of specimen cross-section achieving in a spatial resolution of 12.5 nm between pixels and using a dwell time of 4000 ms/pixel. The "wt% composition vs distance" plots for the different elements (Ge, O) were generated by averaging about 70 horizontal pixel lines across the cross-section area of interest with the help of 'Unary Processing' in TIA software using the quantification outputs from the individual spectra collected at every pixel. Cross-sectional TEM (XTEM) specimens were prepared by standard process of sandwiching, cross-sectioning, grinding, dimple grinding and ion milling with 4 keV Ar ion.

## 2.2. Oxygen Ion implantation

The implantation of the Ge wafers was carried out in the Low Energy Ion Beam Facility (LEIBF) of Inter University Accelerator Centre (IUAC), New Delhi. The Oxygen ions from Electron Cyclotron Resonance Ion Source (ECRIS) were further accelerated to 200 keV, analyzed by high resolution magnetic dipole, focused on the sample by electrostatic quadrupole triplets and scanned for uniform implantation on the sample. The ion beam parameters were obtained from SRIM simulations, whose results summarized in Figure 2. We have used the following equation to fit the simulation data to the Gaussian:

$$n(x) = n_0 exp\left[\frac{-(x-R_p)^2}{2\sigma_p^2}\right] \qquad (1)$$

Where normalization factor $n_o = \frac{Q_T}{\sigma_p\sqrt{2\pi}}$, $Q_T$ is the total number of implanted ions-cm$^{-2}$

At this energy the implanted ions show a projected range of $R_p \approx 328$ nm and the straggle $\sigma_p \approx 99$ nm. The simulation was done for an amorphous target. However, given the crystalline and oriented nature of the Ge wafer, the implantation was done in a high vacuum chamber at room temperature avoiding channelling by vertical tilting the sample by 5 to 7 degrees. Ge substrate was mounted with the help of conducting adhesive on a copper target ladder to avoid sample heating during implantation. The implantation was carried out with average beam current of 10 pµA (particle micro Ampere) for a duration of about 8 hours and 30 minutes. This leads to implanted fluence of $1.9 \times 10^{18}$ ions-cm$^{-2}$.

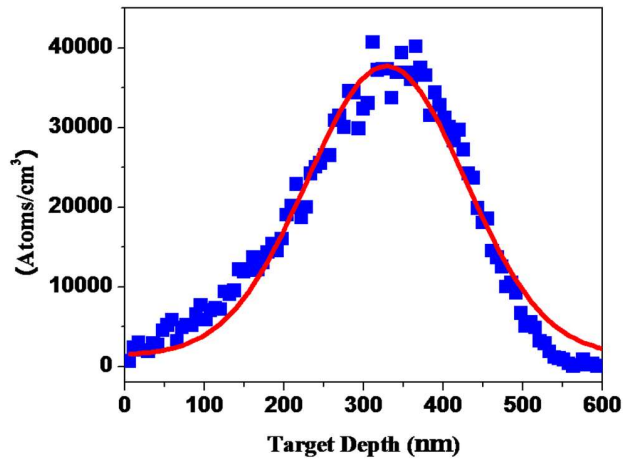

Fig. 2. Plot of Oxygen atom distribution profile Vs Depth as obtained from the SRIM simulation. The solid line is the fit with a Gaussian with projected range 328 nm and straggle 99 nm.

3. Results and discussion

The analysis of data from various characterizations as discussed below, show that in the process of Oxygen ion implantation, that creates the embedded $GeO_2$, the top layer of Ge is amorphized along with a low density nanocrystalline phase. The process of RTA restores the crystallinity of the embedded $GeO_2$ as well as that of the top Ge layer. This establishes that a GeOI can be fabricated from Ge wafer using energetic Oxygen ion implantation that has a finite thickness of GeO$_2$ as BOX and a top layer of crystalline Ge. The results from the various characterization techniques are discussed below.

### 3.1. Raman Spectroscopy

Ge single crystal has a characteristic Raman line at 300cm$^{-1}$ arising from the degenerate LO/TO mode [27]. This line has been widely used to study different structural modifications in Ge viz., amorphization, pressure induced changes etc. [28]. We have also used the particular Raman line to track the changes that occur in the wafer after implantation andupon subsequent RTA. Given the limited penetration depth of the LASER excitation that also depends on the excitation wavelength, one can effectively study the top Ge layer using Raman Spectroscopy. In Figure 3, we show the Raman spectra (excitation 488nm) of the starting Ge wafer along with those of the as- irradiated sample and the irradiated sample after RTA.

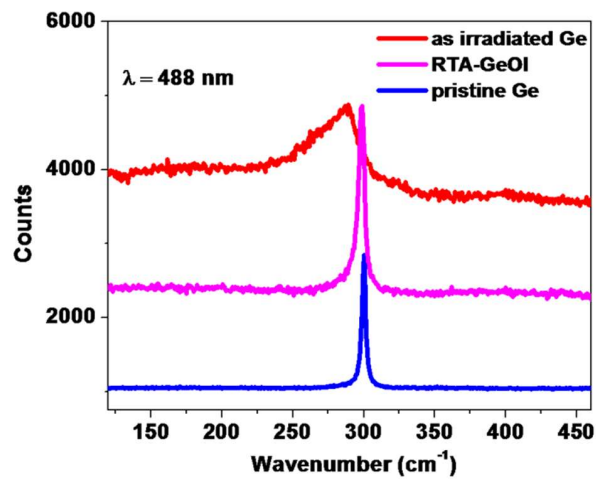

Fig. 3. A comparison of Raman spectra of the samples at different stages of process with that of the pristine Ge wafer.

The figure establishes the basic result that the broad (and shifted) Ramanspectrum of the as-irradiated sample, coming from the amorphous top layer, is restored by the RTA to its pristine value. Figure 3 shows the spectrum over an extended range from 120cm$^{-1}$ to 460 cm$^{-1}$. It can be seen that no peak from GeO$_2$ can be detected indicating the absence of any substantial amount of Oxygen within the top Ge layer. GeO$_2$ peaks occur at 113 cm$^{-1}$, 168 cm$^{-1}$ and 444 cm$^{-1}$.

In Figure 4 we show detailed spectra for both the as-irradiated sampleand the same sample after RTA. The spectra have been taken with two excitations at 488nm and 785nm with penetration depths of ~19nm and ~50 nm respectively [29]. This allows depth resolved information. For the as- irradiated sample, the Raman spectrum at 488 nm, that gets information from the top layer ~ 19nm, shows a skewed nature. The spectrum is analyzed with two Lorentzian peaks (fit parameters in Table I). The spectrum arises from a low density amorphous Ge (Raman line - peak 1) at the top layer of the as-irradiated sample. This matches well with Raman spectra reported for amorphous Ge [28,30,31]. The other component (Raman line- peak 2) is likely to be a co-existing phase of tensile strained nanocrystalline Ge. Such shifts of Ge TO mode to lower frequency have been observed in Ge membranes where the 300cm$^{-1}$ Raman line shifts to lower frequency due to large tensile strain [32-34]. For the spectrum taken at longer wavelength of 785 nm the information has been obtained from a greater depth. As the depth increases the prominence of the strained nanocrystalline component (peak 2) increases compared to the amorphous phase in the top layer. The relative intensity of the peak 2 (see table) is enhanced by a factor of nearly 2. The Raman line for the amorphous part does not shift much (~273cm$^{-1}$) but the nanocrystalline phase at the greater depth shows strain relaxation as the Raman line shifts significantly to higher value of 292 cm$^{-1}$.There is a substantial reduction (by a factor of 2) of the width of the Raman line signifying a decrease in disorder as well as with the tensile strain as the depth increases.

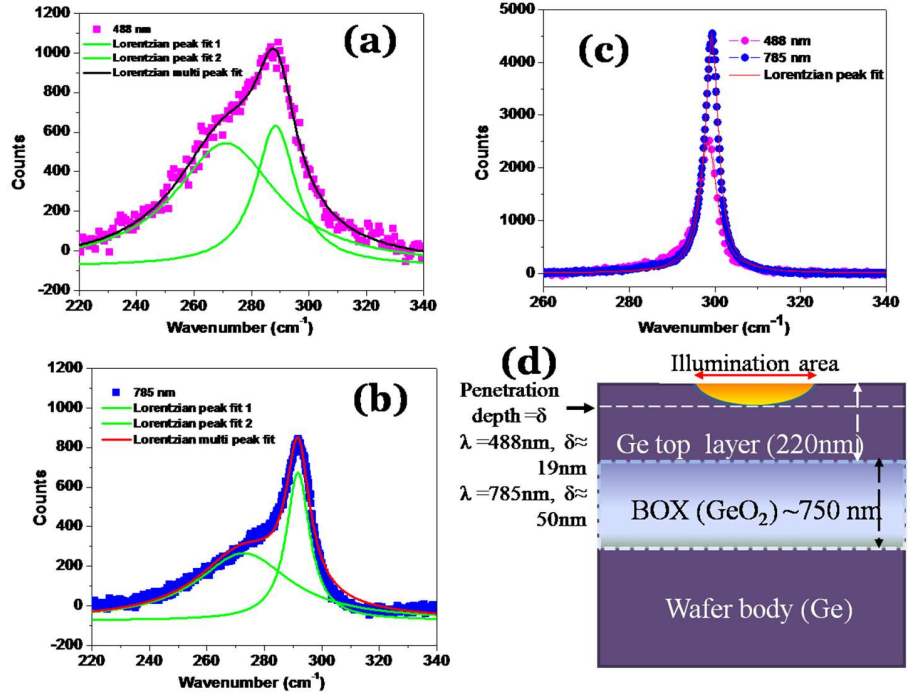

Fig. 4. Depth dependent Raman spectroscopy with two LASER excitations. On as-irradiated Ge wafer for excitation (a) 488 nm and (b) 785nm (c) Irradiated Ge wafer after RTA (d) Schematic of the penetration depth for different wavelength of the exciting Laser.

The RTA of the as-irradiated sample leads to recrystallization of the top Ge layer and the Raman Spectrum shows that the peak shifts closer to the pristine value of 300 cm$^{-1}$. From Table I it can be seen that there is a small residual shift, $\Delta\omega = \omega_{observed} - \omega_{pristine}$. This can be used to find the bi-axial strain $\varepsilon_\parallel$ in the Ge layer using the relation $\Delta\omega = b\varepsilon_\parallel$ where $b = -(415 \pm 40)$cm$^{-1}$ [34]. The negative sign indicates a tensile strain $\varepsilon_\parallel > 0$ that leads to a downward shift of the Raman line $\Delta\omega < 0$ [32,33]. From Table I the strain was evaluated for the excitation of 488nm for the shift $\Delta\omega = 1.64$ cm$^{-1}$ and the strain was estimated to be $\varepsilon_\parallel \approx +0.4\%$. In case of excitation with 785 nm the shift is $\Delta\omega = 0.74$ cm$^{-1}$ and $\varepsilon_\parallel \approx +0.18\%$. The homogenous strain in the top Ge layer (that stretches the lattice) thus gets reduced as we probe greater depth of the layer. The extent of disorder or inhomogeneous strain given by the width of the spectrum also gets reduced upon annealing and it also with increased depth in the layer. The Raman data also shows that at least to the depth probed by the Raman spectroscopy (~50nm) there is no signature of GeO$_2$ line indicating absence of significant oxygen contamination at the top layer.

Table I. Depth dependent Raman spectroscopy in as-irradiated Ge wafer and after RTA. The table also shows the biaxial strain $\varepsilon_\parallel$ in the annealed top Ge layer calculated from the shift in peak position from pristine wafer

| Excitation Wavelength (nm) | As-irradiated sample (cm$^{-1}$) | | | | After RTA (cm$^{-1}$) | |
|---|---|---|---|---|---|---|
| | Peak no | Peak position | Intensity ratio | Width | Peak position | Width |
| 488 | 1 | 271.11 | $I_2/I_1=$ 1.14 | 43.69 | 298.36 | 5.07 |
| | 2 | 288.42 | | 17.06 | $\varepsilon_\parallel \approx +0.4\%$. | |
| 785 | 1 | 272.85 | $I_2/I_1=$ 2.20 | 39.43 | 299.26 | 3.80 |
| | 2 | 291.66 | | 9.92 | $\varepsilon_\parallel \approx +0.18\%$. | |

*3.2. Transmission Electron Microscopy*

While the Raman spectra provide information on the top Ge layer and restoration of its crystallinity by the RTA process, a detailed investigation using analytical TEM of cross-sectioned layer provides theinformation on the formation of the buried oxide layer, its chemical constitution and also, the nature of the interface so formed due to the implantation.

In Figure 5(a) the high-angle annular dark-field scanning transmission electron microscopy (HAADF-STEM) image shows the cross-section of the layered structure in the final RTA annealed GeOI sample. The buried oxide layer of finite thickness and the top Ge layer can be clearly identified in this image. The STEM-EDS line-averaged concentration profiles for Ge and O across the layer thickness in the cross-sectioned sample after RTA are presented in Fig. 5(b) and (c) respectively.

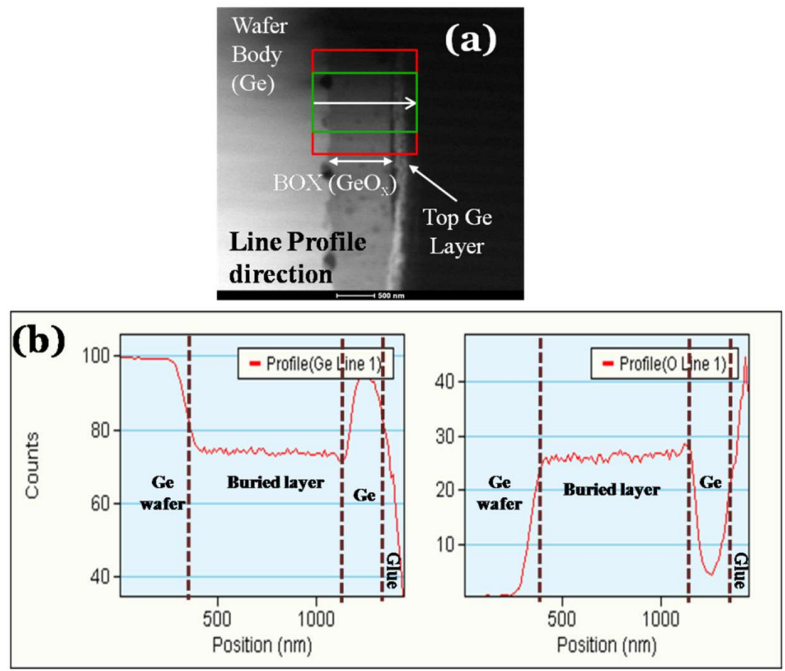

Fig. 5. (a) STEM-HAADF image of the cross-sectioned RTA annealed sample revealing layered structure. (b) STEM-EDS line-averaged concentration (wt.%) profiles of Ge-K and O-K.

Both these line-profiles were obtained by integrating 70 horizontal line profiles (within green box in Fig.5(a) across the thickness of the layer cross-section, after acquiring the pixel-wise quantitative STEM-EDS mapping, acquired by 'spectrum-imaging' technique, over a suitable area of interest (as marked by red box in Fig.5a) making sure that the STEM e-beam scanning in the x-direction is perpendicular to the layer-interfaces exposed in the cross-sectioned sample. Integration of 70 horizontal line-profiles has generated a composition profile with much reduced noise. One important observation can be made that the distribution of oxygen in the buried layers follow a 'top-hat' type distribution and is nearly flat within an identified layer unlike the Gaussian distribution expected from Figure 2. The implanted oxygen gets redistributed due to transient enhanced diffusion during implantation and subsequent redistribution during annealing.

Table II. Mean thicknesses of buriedoxide layer, Top Ge layer and the interfaces

| Layer | Post RTA (nm) |
|---|---|
| Buried Implanted | 740 |
| Top Ge | 220 |
| Interface 1 (Bulk Ge-Buried layer) | 85 |
| Interface 2 (Buried-Top Ge) | 85 |

Uncertainty in thickness measurements ≈ ± 10nm

The buried oxide layer that is created by the implantation has an average thickness of 760±10nm with the centre of the layer situated at a distance of 560±10 nm from top. The top Ge layer is clearly identified and has substantial thickness of nearly 200nm. The uncertainty in layer thickness mainly arises because the top layer has an undulated structure as shown in the Scanning Probe Microscopy image given in Figure S1of the supplementary data. The mean position and thickness of the implanted layer is greater than 328 nm that predicted by the SRIM simulation. It may be due to the crystallinity of the sample and the transient processes during implantation and successive redistribution during RTA.

The embedded oxide layer has two interfaces. One is with the bulk of the Ge wafer and the other is the top Ge layer. The interface between the buried layer and the bulk Ge is diffused. There is diffusion of Oxygen from the implanted layer to the top layer. However, very close to the top of the Ge layer there is very little Oxygen. This supports absence of $GeO_x$ in Raman spectrum.

From the TEM-EDS line profile data (Figure 5), it is possible to obtain a semi-quantitative estimation of the relative Ge and the Oxygen concentration. This can be used to evaluate the approximate effective composition of the embedded germanium oxide ($GeO_x$) layer. In Table III the effective Ge and O concentrations (in wt %) as obtained from the EDS data are given. The RTA leads to enhancement of the O concentration in the buried oxide layer significantly and the resulting $GeO_x$ has the stoichiometry with $x \approx 1.7$. This is not the expected perfect stoichiometry of $GeO_2$ but is close to it. The exact composition has been discussed later on.

Table III. Ge and O concentration in the buried oxide layer

| Sample | Ge (wt%) | O (wt%) | Wt% ratio (Ge/O) |
|---|---|---|---|
| Post RTA anneal | 73 | 27 | 2.7 |

Uncertainty in concentration is ± 3%

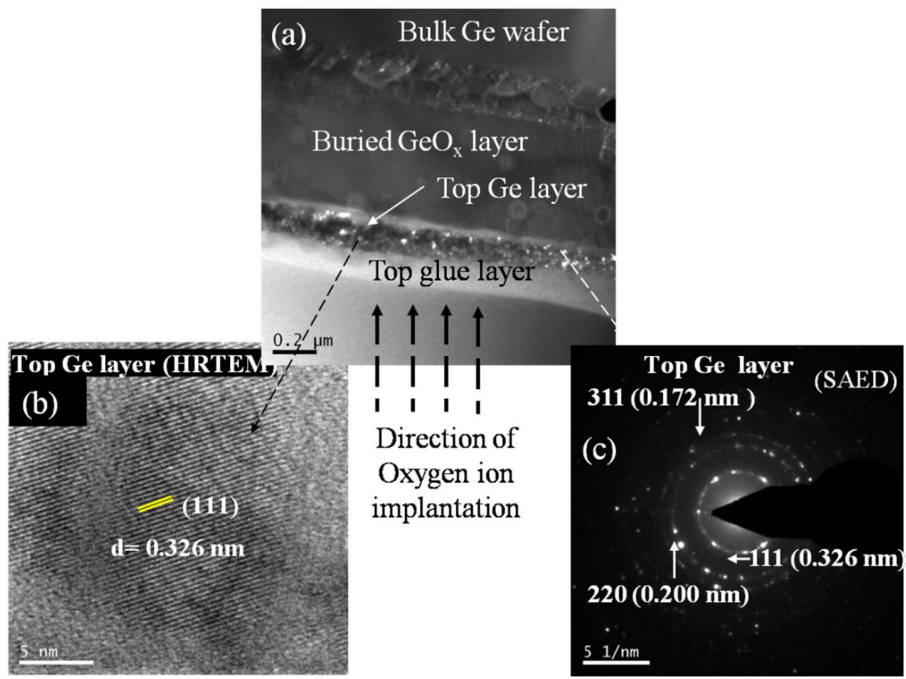

Fig. 6. TEM micrographs of RTA processed cross-section sample. (a) Dark Field TEM Image, (b) HRTEM Lattice image and (c) Corresponding SAED Pattern.

Figure 6 represents detailed TEM Imaging for cross-section of the RTA processed sample. Fig. 6(a) is the TEM Dark Field Image that effectively represents the three layers, namely the top Ge layer, the buried oxide layer and bulk Ge wafer, with distinctive phase separated contrast. The HRTEM lattice imaging (Fig. 6(b)), and Selected Area Electron diffraction (Fig. 6(c)) are the correlated data for qualifying Fig. 6(a). The nature of the quality of the top Ge layer which eventually can be utilized for applications is analyzed. The HRTEM image as well as SAED pattern of top Ge after RTA shows crystallized layer. From the HRTEM data as well as from the diffraction pattern, the lattice spacing corresponding to the Miller indices were used to find the lattice constant of the FCC Ge as $a \approx 0.5669$ nm which corresponds very well with the standard value of $a \approx 0.5658$ nm for crystalline Ge [35]. The small difference between the standard value of $a$ and that of the irradiated and RTA processed sample can be interpreted as a small strain $\approx +0.397\%$. The positive strain implies a tensile strain.

The proper densification of the top Ge layer to its crystalline state corroborates with the observation of the Raman data (Figure 4). In the as-irradiated sample the top layer is amorphous Ge co-existing with strained crystalline Ge with pockets of $GeO_2$ as seen in the Raman data as well as TEM data. The TEM data are shown in Figure S2.

TEM micrographs of top Ge layer in post RTA sample are shown in Figure 7. Figure 7(a) is an image of the multi-facetted nanocrystals, grown on the top layer post-RTA sample. Figure 7(b) is a closer view of facetted nanocrystal. Figure 7(c) is the HRTEM image of a multi-facetted nanocrystal showing facial lattice planes. The detailed TEM investigation thus shows that post-RTA there is indeed a recrystallized Ge layer (total thickness ~220nm) with some nanocrystals. The layer has a diffused boundary with the embedded oxide layer of thickness ~85 nm that contains some Oxygen. However, there is a clean layer of nearly 100nm of Ge that can be used for making device.

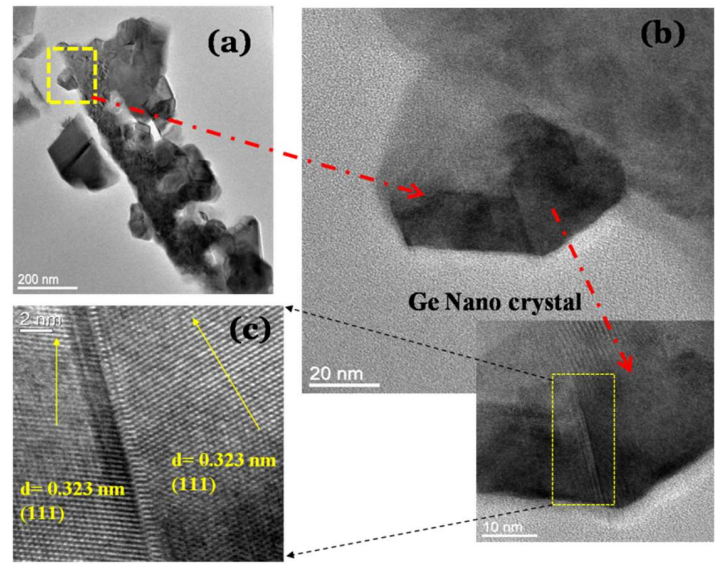

Fig. 7. TEM micrographs of top Ge layer in post RTA sample. (a) Multi-facetted nanocrystals, (b) Closer view of Multi-facetted nanocrystal and (c) HRTEM image of facial lattice planes.

To demonstrate the importance and necessity of short duration RTA, we carried out a prolonged annealing schedule on a sample of as- irradiated Ge in inert (Nitrogen) atmosphere at 650°C for 1 hour. The prolonged anneal does improve the crystalline quality of the buried layer and also that of the top Ge layer but due to diffusion of Oxygen the top Ge layer gets oxidized and has been converted into crystalline $GeO_2$. This can be seen from TEM micrograph shown in Figure 8. Figure 8(a) is the Lattice image (HRTEM) while Fig. 8(b) exhibits the Nano Beam Electron Diffraction (NBED) pattern, both qualifying the crystalline signature of the top $GeO_2$ layer after prolonged time of annealing. NBED technique with a probe size of 1 nm was employed to confirm the nanocrystalline specificity of the referred layer. The Figure 8 (c) also shows the Raman Spectrum of the sample along with TEM data.

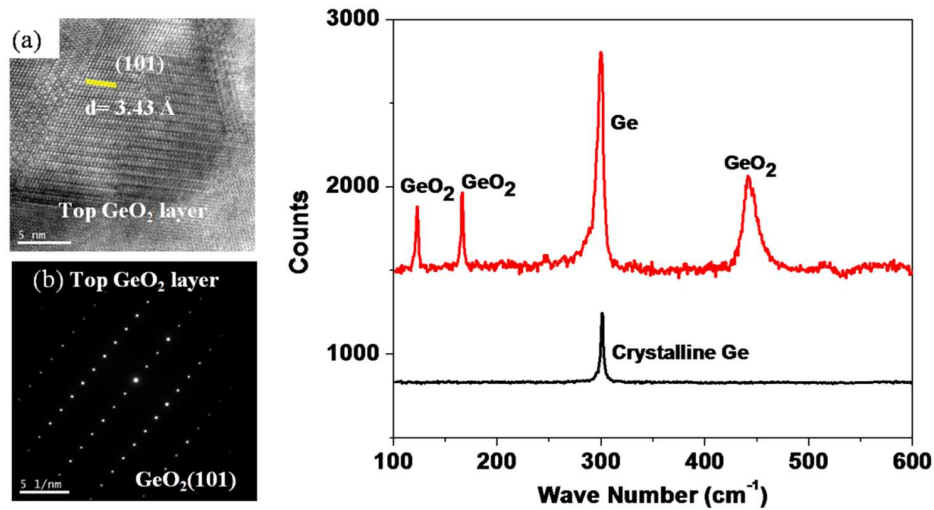

Fig. 8. 8(a) HRTEM micrograph showing Lattice image, 8(b) NBED pattern of the top $GeO_2$ layer and 8(c) Raman Spectra of the long time annealed sample along with the spectrum of crystalline Ge Wafer.

It can be seen that the Raman spectrum of the sample with prolonged annealing shows peaks arising from $GeO_2$ which is not seen in the sample with RTA process (Figure 3). The sharp line due to Ge in the Raman spectrum arises from pockets of Ge present in the $GeO_2$ layer after prolonged anneal.

This establishes that post irradiation a short duration RTA restores the crystallinity in the top Ge layer in the as-irradiated sample and does not lead to oxidation of the top Ge layer which can be used for device applications. The Oxygen irradiated and RTA annealed sample thus can be referred to as the Germanium- on -Insulator (GeOI) sample.

*3.3. XRD and Rocking curve*

Figure 9 shows the wide angle XRD taken on the final RTA annealed sample (GeOI). The data shows 2 peaks arising from crystalline Ge identified with cubic crystal indexing of (220) and (400). We have used the peak at (400) to obtain the rocking curve (vary $\omega$ at fixed $2\theta$). This is shown in Figure 10. From the XRD data we calculated the lattice constant for the (400) peak in the GeOI sample and compared to standard Ge data (Hom et.al (1975)). The observed strain is tensile +0.488%. This is comparable to ≈ +0.397% obtained from TEM data. The observed strain is more than the uncertainty in strain data of TEM ≈ ±0.1%. This is similar to the strain seen in the Raman Spectroscopy (Table I).

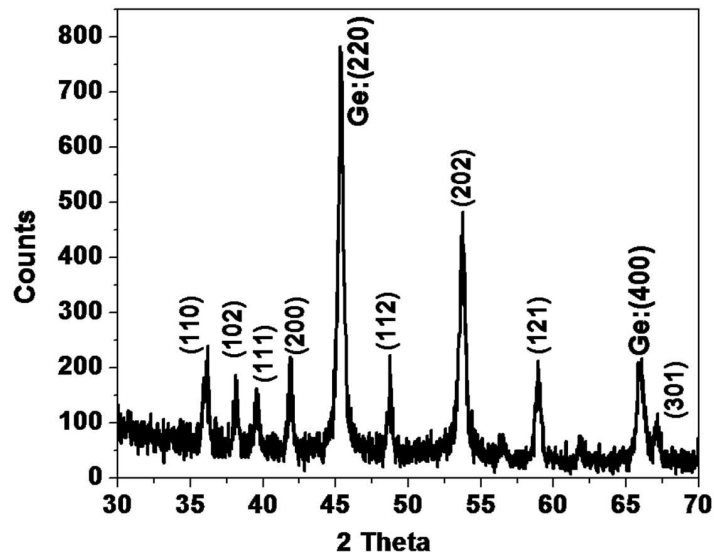

Fig. 9. Standard X-Ray diffraction curve taken on ion implanted and RTA processed Ge wafer. The data shows peaks for Ge as well as from the embedded oxide layer. The two peaks for Ge belong to Ge (220) and (400). Other peaks have been indexed with crystalline Hexagonal $GeO_2$.

The strain calculated above is the homogeneous tensile strain that arises from the expansion of lattice parameter after irradiation and re-crystallization due to RTA. To obtain an estimate of the dislocation density we used the FWHM from the rocking curve data shown in Figure 10. Since no filter was used for CuK$_\alpha$ we fitted the observed XRD peak with two Lorentzians to get the FWHM (ΔΩ).

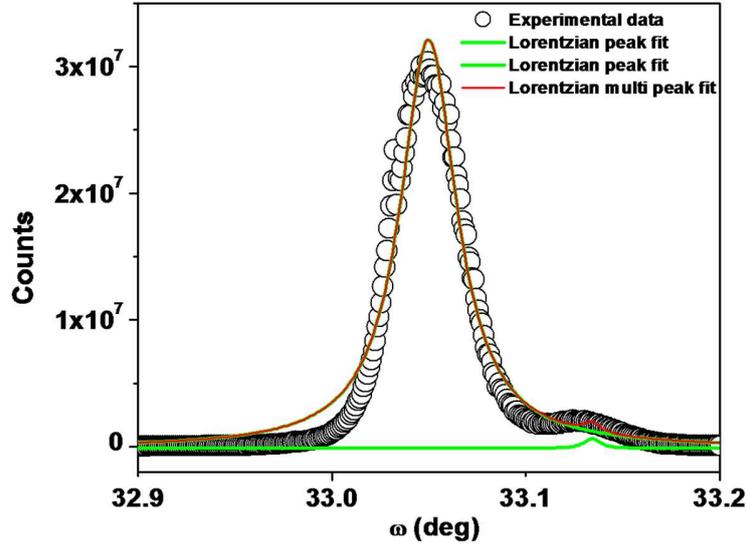

Fig. 10. Rocking curve of the irradiated wafer after RTA process. The small peak at 33.13$^0$ is due to the CuKα2 line.

From the FWHM the inhomogeneous strain is estimated to be 0.127% ($=\frac{\Delta\Omega}{\Omega}$). The width (FWHM) of the rocking curve has been used to calculate the slip dislocation density using Hirsch model [36,37]

Dislocation density = $\left(\frac{\hat{\beta}^2}{9\hat{b}^2}\right)$ (2)

where $\hat{\beta}$ is the FWHM of the rocking curve in radians and $\hat{b}$ is the Burger vector in cm. For Ge $\hat{b}$ =3.99x10$^{-8}$ cm [38]. The value of dislocation density was found to be 2.67x10$^7$ cm$^{-2}$ (For the pristine wafer from the rocking curve the dislocation density is ≈1x10$^6$ cm$^{-2}$ as calculated from the Rocking curve, data not shown). This dislocation density value is an over-estimation because it was assumed that the FWHM value is contributed solely by the dislocation. There are other factors that also contribute to $\hat{\beta}$ [37] and they have not been factored out. In Ge grown on matched substrate the dislocation density can be around $10^7$ cm$^{-2}$. Using different post growth processes the dislocation density can be brought down to $\geq 10^6$ cm$^{-2}$ [39,40]. Thus the observed dislocation density though not very low as in the pristine wafer, however, is within a reasonable limit.

From the XRD data shown in Figure 9, that show the GeO$_2$ peaks from the embedded layer one could analyze the quality of the GeO$_x$ layer. In Table IV below we compare the lattice constants determined from the XRD data with the standard data ICDD# 36-1436 [41]. The structure is Hexagonal.

Table IV Lattice constants of the GeO$_2$ in the BOX

|  | a (Å) | b (Å) | c (Å) |
|---|---|---|---|
| JCPDS File | 4.985 | 4.985 | 5.648 |
| Experimental | 4.969 | 4.969 | 5.586 |

The data show that the oxide in the BOX is a good quality GeO$_2$. It has small compression of $a$ and $b$ axes with compressive strain $\frac{\Delta b}{a} = \frac{\Delta a}{a} \approx \frac{a_{expt} - a_{JCPDS}}{a_{JCPDS}}$ = -0.32% (compressive). The $c$ axis also has a compression with strain $\frac{\Delta c}{c} \approx \frac{c_{expt} - c_{JCPDS}}{c_{JCPDS}}$ = -1.1% (compressive). However, the hexagon is undistorted because the observed $\frac{c}{a}$ ratio is 1.124 very close to that published the JCPDS file (1.132).

No detailed electrical measurements on the GeOI was made that would enable us to assess the electrical quality of the wafer. However, preliminary measurement of resistance and Hall mobility in the final annealed sample shows a reduction of mobility by a factor of 1.5 to 775

cm$^2$/V-s. The top layer in plane resistivity of 32 Ohm-cm is comparable to but somewhat less than 44 Ohm-cm for the pristine wafer. But the out of plane resistivity which involves transport through the BOX approaches 90 Ohm-cm which is larger than the resistivity of the pristine wafer by a factor of more than 2.

## 4. Conclusions

In summary, the Germanium-on-insulator (GeOI) structure in Germanium wafer has been developed using 200 keV Oxygen ion implantation at a fluence of 1.9 x 10$^{18}$ ions-cm$^{-2}$ followed by rapid thermal annealing at 650$^0$C. The top Ge layer of thickness of ~ 220 nm is produced on BOX layer of crystalline GeO$_2$ having thickness of about 0.75μm. Raman Spectroscopy, Cross-sectional HRTEM with SAED and EDS analyses have been carried out to establish that the top Ge layer is recrystallized during annealing with faceted crystallites. The top layer has a small tensile strain of around +0.4% and has estimated dislocation density of 2.7X10$^7$cm$^{-2}$. The thickness of the top layer and its resistivity is such that it can be utilized for device fabrication.

**Acknowledgement**

AKR acknowledges financial support from Science and Engineering Research Board (SERB), Government of India as a SERB Distinguished Fellow (project number: SB/DF/008/2019). AG acknowledges financial support from Technical Research Centre Project funded by the Department of Science and Technology, Government of India.

## References

[1] A.Ploeul, G. Kraeuter, Solid-State Electron. 44 (2000) 775–782.

[2] M. W.Geis et al, Optics Express 15 (2007) 16888–95.

[3] R. Huang, Q. Huang, S. Chen, C. Wu, J. Wang, X. An, Y.Wang, Nanotechnology 25 (2014) 505201.

[4] D. P. Tran, J. Macdonald Thomas, B. Wolfrum, R. Stockmann, T. Nann, A. Offenhausser, B Thierry, Appl. Phys. Lett. 105 (2014) 231116.

[5] S. Sett, V. K. Aggarwal, A. Singha, S. Bysakh and A. K. Raychaudhuri, Semicond. Sci. Technol. 35(2020) 025020.

[6] T. Akatsu, C. Deguet, L. Sanchez, F. Allibert, D. Rouchon, T. Signamarcheix, C. Richtarch, A. Boussagol, V. Loup, F. Mazen, J. M. Hartmann, Y. Campidelli, L. Clavelier, F. Letertre, N. Kernevez, C. Mazure,Materials Science in Semiconductor Processing 9 (2006) 444.

[7] Delphine Marris-Morini et al, Nanophotonics7(11) (2018) 1781-1793.

[8] M. El Kurdi, S. David, X. Checoury, G. Fishman, P. Boucaud, O. Kermarrec, D. Bensahel, B. Ghyselen, Optics Communications 281 (2008) 846-850.

[9] Z. Zheng, X. Yu, M. Xie, R. Cheng, R. Zhang, and Y. Zhao, Appl. Phys. Lett. 109 (2016) 023503.

[10] C. H. Lee, T. Nishimura, T. Tabata, D. D. Zhao, K. Nagashio, A. Toriumi, Appl. Phys. Lett. 102 (2013) 232107.

[11] Wen-Hsin Chang, Toshifumi Irisawa, Hiroyuki Ishii, Hiroyuki Hattori, Hideki Takagi, Yuichi Kurashima, Tatsuro Maeda, Appl. Phys. Express 9 (2016) 091302.

[12] S. Nakaharai, T. Tezuka, N. Sugiyama, Y. Moriyama, S. Takagi, Appl. Phys. Lett. 83 (2003) 35.

[13] Y. Liu, M. D. Deal, J. D. Plummer, Appl. Phys. Lett. 84 (2004) 2563.

[14] Kwang Hong Lee,Shuyu Bao, Gang Yih Chong, Yew Heng Tan, Eugene A. Fitzgerald, Chuan Seng Tan, J. Appl. Phys. 116 (2014) 103506.

[15] Kwang Hong Lee,Shuyu Bao, Gang Yih Chong, Yew Heng Tan, Eugene A. Fitzgerald, Chuan Seng Tan, APL Mater. 3 (2015) 016102.

[16] I. P. Ferain, K. Y. Byun, C. A. Colinge, S. Brightup, M. S. Goorsky, J. of Appl. Phys. 107 (2010) 054315.

[17] Munho Kim, Sang June Cho, Yash Jayeshbhai Dave, HongyiMi, Solomon Mikael, Jung-Hun Seo, Jung U Yoon, Zhenqiang Ma, Semicond. Sci. Technol. 33 (2018) 015017.

[18] Xiao Yu, Jian Kang, Rui Zhang, Mitsuru Takenaka, Shinichi Takagi, Solid-State Electronics 115 (2016) 120-125.

[19] U. Gosele et al, Microelectronic Engg. 28 (1995) 391.

[20] F Balestra, A Nazarov, D. Lysenko, NATO Science Series II Springer 58 (2002).

[21] Miao Zhanga, Zhenghua Ana, Chenglu Lina, Paul K. Chub, Mat. Sci and Engg. B 114 (2004) 115255.

[22] Zhenghua An, M. Zhang, Ricky K.Y. Fu, Paul K. Chu,Chenglu Lin, J. of Electronics Materials 33 (2004) 207.

[23] H.J. Stein, J. Electrochem. Soc. Solid State Sci. and Tech.121 (1974) 1073.

[24] N.M. Ravindra et al, Nuclear Instruments and Methods in Phy. Research B 46 (1990) 409-412.

[25] Qi-Chu Zhang, J. C. Kelly, M. J. Kenny,Nuclear Instruments and Methods in Phy. ResearchB47 (1990) 257-262.

[26] S.R. Elliot, Physics of Amorphous Materials, Wiley, New York (1989).

[27] J. H. Parker Jr., D. W. Feldman and M.Ashkin, Phys. Rev. 155 (3) (1967) 712.

[28] F. Coppari,J. C. Chervin, A.Congeduti, M. Lazzeri, A. Polian, E. Principi, A. Di Cicco, Phys. Rev. B 80 (2009) 115213.

[29] https: //www.horiba.com/fileadmin/uploads/Scientific/Documents/Raman/Semiconductors01.pdf

[30] R. Alben, J. E. Smith, M. H. Brodsky, D. Weaire, Phys. Rev. Lett. 30 (1973) 1141.

[31] P. V. Santos, L. Ley, Phys Rev B 36 (1987) 3325.

[32] Y. Y. Fang, J. Tolle, R. Roucka, A. V. G. Chizmeshya, John Kouvetakis, V. R. D'Costa, José Menéndez, Appl. Phys. Lett. 90 (2007) 061915.

[33] Yijie Huo, Hai Lin, Robert Chen, Maria Makarova, Yiwen Rong, Mingyang Li, Theodore I. Kamins, Jelena Vuckovic, James S. Harris, Appl. Phys. Lett. 98, (2011) 011111.

[34] Yize Stephanie Li, John Nguyen, Scientific Reports 8 (2018) 16734.

[35] T. Hom, W. Kiszenik, B. Post, J. Appl. Cryst. 8 (1975) 457.

[36] P. Gay, P. B. Hirsch, A. Kelly, Acta Met. 1 (1953) 315.

[37] J. E. Ayers, Journal of Crystal Growth 135 (1994) 71-77.

[38] https://engineering.dartmouth.edu/defmech/chapter_9.htm

[39] Hsin-Chiao Luan, Desmond R. Lim, Kevin K. Lee, Kevin M. Chen, Jessica G. Sandland, Kazumi Wada, Lionel C. Kimerling, Appl. Phys. Lett. 75 (1999) 2909.

[40] Kwang Hong Lee, ShuyuBao, Gang Yih Chong, Yew Heng Tan, Eugene A. Fitzgerald, Chuan Seng Tan, APL Mater. 3 (2015) 016102.

[41] Haiping Jia, Richard Kloepsch, Xin He, Juan Pablo Badillo, Martin Winter, Tobias Placke, J. Mater. Chem. A 2 (2014) 17545.